\documentclass{PoS}

\title{Pulse spectral evolution of GRBs: implication as standard candle}

\ShortTitle{Pulse spectral evolution of GRBs}

\author{\speaker{Rupal Basak}\\
        Tata Institute of Fundamental Research, Mumbai\\
        E-mail: \email{rupalb@tifr.res.in}}

\author{A. R. Rao\\
        Tata Institute of Fundamental Research, Mumbai\\
        E-mail: \email{arrao@tifr.res.in}}

\abstract{Using an \emph{empirical} description of a prompt GRB pulse, we analyze the individual pulses of all Fermi/GBM GRBs 
with known redshifts, till July 2009. This description is simultaneous in time and energy and allows one to determine 
the peak energy of Band spectrum at zero fluence ($E_{peak,0}$). We demonstrate, for the first time, that the $E_{peak,0}$ bears a 
very strong correlation with the isotropic energy of the individual pulses, and hence, each pulse can be used as a luminosity indicator. 
As a physical description is needed in order to use GRB pulses for cosmological purposes, we explore other physical spectral models. As pulses 
are the building blocks of a GRB, we choose another sample of Fermi/GBM GRBs having bright, long and single/ separable pulse(s) and fit the 
time-resolved spectra of the individual pulses with the Band model and a model consisting of a blackbody and a power-law. Both these models 
give acceptable fits. We find that the peak energy/ temperature always decreases exponentially with fluence in the later part of a pulse. 
We investigate multiple spectral components in the initial rising part and provide a comprehensive empirical description of the 
spectral and timing behaviour of prompt GRB pulses. This work strongly extends the possibility of using GRB pulses as standard candles 
and the spectral parameters as proxy for redshift.}

\FullConference{Gamma-Ray Bursts 2012 Conference -GRB2012,\\
		May 07-11, 2012\\
		Munich, Germany}

\begin{document}

\section{Introduction}

%Band's GRB model (Band et al. 1993) has a very good agreement with a large set of data 
The spectral model suggested by Band et al. (1993) gives very good fits to the GRB spectra
obtained 
from a variety of instruments, but this model is difficult to reconcile with any
physical scenario, like synchrotron emission.  In recent times, researchers have also reported multiple spectral 
components while fitting the prompt emission spectra in a wider band (Zhang et al. 2011) --- the Band model is not sufficient to capture
the whole electromagnetic spectrum. Ryde (2004) have shown that instantaneous spectra of GRB pulses sometimes disagree with the
Band model, but can be described by a combination of a blackbody and a power-law (BBPL). The temperature of the
BBPL model evolves with time. Ryde et al. (2010) have also  found that the time resolved spectra of GRB 090902B do not agree even
with the BBPL model, but a multi-colour BB with a power-law fits the data. Hence, no unified picture of GRB prompt spectrum has emerged yet.

It is very important to obtain a correct description of the GRB spectrum, not only to understand the emission mechanism, but,
a comprehensive physical description is essential to use GRBs for cosmological purposes. Basak \& Rao (2012a) have shown
that the individual \emph{pulses} of a GRB can be parametrized by a set of variables, using an empirical law of hard-to-soft
spectral evolution. Basak \& Rao (2012b) have used this model to find peak energy at the very beginning of the pulses ($E_{peak,0}$)
of a set of Fermi/GBM GRBs with measured redshifts. They have shown that $E_{peak,0}$ shows a very strong correlation 
with the isotropic energy ($E_{\gamma,iso}$) of the pulses. This shows that the individual pulses of a GRB can be separately treated,
and each of them can be used as a standard candle, instead of the full GRB.

The pulse description in Basak \& Rao (2012b), however, is an empirical one. It is worthwhile to find whether other physical model(s)
is (are) consistent with the data. The realization that the pulses are the building blocks of a GRB also motivates one to try
various models and spectral evolution in a single pulse. The strategy is to establish a comprehensive description for a 
single pulse, then use them for more complicated pulses of GRBs (having known redshifts), and use these individual pulses
as standard candles in cosmology. We collect such a set of 11 GRBs from the catalogue of Nava et al. (2011). In this paper
we present the main conclusions of the early works (Basak \& Rao 2012a; b) and then focus on the new findings.
The detailed results will be published elsewhere (Basak \& Rao, in preparation; Rao et al. ApJ submitted).
The final aim is to use these comprehensive descriptions to parametrize prompt GRB pulses by physical models. 

\section{Analysis and Results}

\subsection{Simultaneous timing and spectral description of a GRB pulse: $I(t,E)$}
The set of GRBs used by Basak \& Rao (2012b) for a correlation study are 9 Fermi/GBM GRBs with known redshifts, till July, 2009. They
used the simultaneous timing and spectral description of the intensity as a function of
time, t, and energy, E, ($I(t,E)$), developed by Basak \& Rao (2012a), to describe individual 
pulses of these GRBs. A concise description of finding $I(t,E)$ is as follows: 

Consider Norris model (Norris et al. 2005) for the light curve, $I(t, ~A_n,~ \tau_1, ~\tau_2, ~t_s)$, integrated over energy. Here, $A_n$ 
is the normalization, $\tau_1$ and $\tau_2$ are the time constants of the rising and falling parts of the pulse, respectively, and $t_s$ is 
the pulse start time. Assume that $\tau_1$, $\tau_2$ do not vary with energy. Consider Band model for the spectrum at each time bin, 
$I(E, ~A_b,~ \alpha, ~\beta, ~E_{peak}~(t,~E_{peak,0},~ \phi_0))$, where $A_b$ is the normalization, $\alpha$ and $\beta$ are the low and
high energy photon indices, respectively, and $E_{peak}$ is the peak energy of the spectrum. Here, $E_{peak,0}$ and $\phi_0$ 
are parameters of $E_{peak}$ evolution (Liang \& Kargatis 1996), given by 
$E_{peak}(t)=E_{peak,0}~exp \left(-\frac{\int_{t_s}^t \! I(t') \, dt'}{\phi_0}\right)$. 
Here we have assumed that $\alpha$ and $\beta$, determined from the time-integrated spectrum, do not vary with time. Assuming a set 
of $E_{peak,0}$ and $\phi_0$ values, Band spectrum can be generated over the time bins, which gives a three dimensional model ($I(t, E)$)
of the pulse. If we assume a grid of $E_{peak,0}$ and $\phi_0$ values, then $I(t, E)$ is determined at each grid point. To find the
correct value of $E_{peak,0}$ and $\phi_0$, we integrate $I(t, E)$ over the time and thus generate a 2 parameter XSPEC table model.
This is used for fitting the time-integrated data by $\chi^2$ minimization technique to find the best fit values of $E_{peak,0}$ and 
$\phi_0$. Once all the parameters are determined, the pulse can be regenerated and
various timing properties (e.g., light curve at various energies, width variation with energy, spectral lag) can be obtained.

\begin{figure}[h]\centering
 \includegraphics[width=3.0 in]{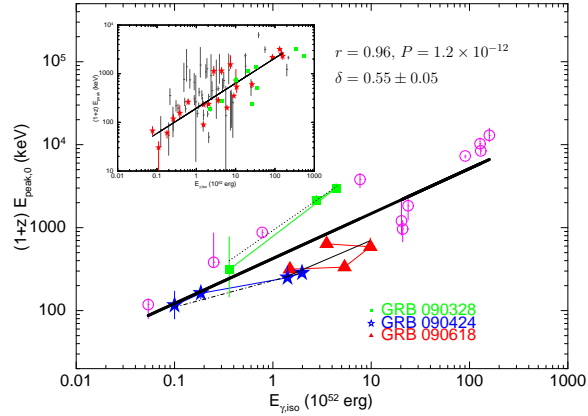} 
\caption{Correlation of isotropic energy ($E_{\gamma,iso}$) with $E_{peak,0}$. The other correlations are shown in the 
inset. See text for explanations.}
\end{figure}

\subsection{An exciting outcome: $E_{peak,0}$-$E_{\gamma,iso}$ correlation}
In the prompt emission, GRBs show a variety of correlation between a spectral or a timing parameter with an energy related 
physical parameter. For example, peak energy of GRB ($E_{peak}$) correlates with the isotropic energy ($E_{\gamma,iso}$), known as the
\emph{Amati correlation} (Amati et al. 2002). This correlation can be studied for a set of GRBs, within the pulses of
the set, or within the time-resolved spectra of the set. The results of our correlation studies are as follows (see Figure 1):

(a)\emph{Time-integrated correlation}: The Amati correlation is rather weak. The Pearson correlation coefficient, r
= 0.80 with chance probability, P = 0.0096 (shown by green boxes in the inset of Figure 1). 

(b) \emph{Time-resolved correlation}: The correlation is lost (r=0.37; see the black dots in the inset of Figure 1). 
Ghirlanda et al. (2010) showed that the correlation of $E_{peak}$ with isotropic peak luminosity ($E_{peak}$-$L_{iso}$),
known as the \emph{Yonetoku correlation} (Yonetoku et al. 2004), holds within the time-resolved spectra of these GRBs. This is 
a very convincing confirmation in favour of the reality of such correlation, and a very strong argument against any instrumental 
bias. The fact that $L_{iso}$ is a better description in a time-resolved spectrum, leads to such a correlation within a GRB. For Amati 
correlation to hold within a GRB, time-resolved spectral data is not the correct choice. 

(c) \emph{Pulse-wise correlation}: Krimm et al. (2009) showed that Amati correlation holds within the broad pulses of GRBs. 
We calculate the $E_{peak}$ of the pulses of the GRBs and examine the Amati correlation. The correlation is not only restored, 
but it improves to r = 0.89 with P = 2.95 $\times 10^{-8}$. These are shown by red stars (Figure 1, inset).

(d) \emph{$E_{peak}$ replaced by $E_{peak,0}$}: The fact that the Amati correlation improves with pulse-wise analysis indicates 
that individual pulses are independent and should be analyzed separately. However, $E_{peak}$ is a 
pulse average quantity and hence lacks the information of the spectral evolution. Hence, we replace $E_{peak}$ with $E_{peak,0}$, 
which is the peak energy a pulse starts with. The scatter plot is shown in Figure 1. The correlation now improves to 
r = 0.96 with P = 1.6 $\times 10^{-12}$. In this plot we have marked different pulses of some GRBs with colours and 
markers to show that the pulses of a GRB follow the same correlation. Data points of these pulses are joined by the corresponding 
coloured lines and correlations are shown by black lines (dotted for GRB 090328, dot-dashed for GRB 090424 and solid line for GRB 090618).
The slope of the straight line fitting the log-log plot is, $\delta=0.55 \pm 0.05$. The Amati correlation gives a corresponding
slope of $E_{peak}$-$E_{\gamma,iso}$ correlation, $\delta=0.52 \pm 0.06$. Hence, it is clear that the slope of the relation 
between the peak energy (either $E_{peak}$ or $E_{peak,0}$) and isotropic energy is similar, though the $E_{peak,0}$ values are
always greater than the average $E_{peak}$.

\begin{figure}[h]\centering
  \includegraphics[width=3.2 in]{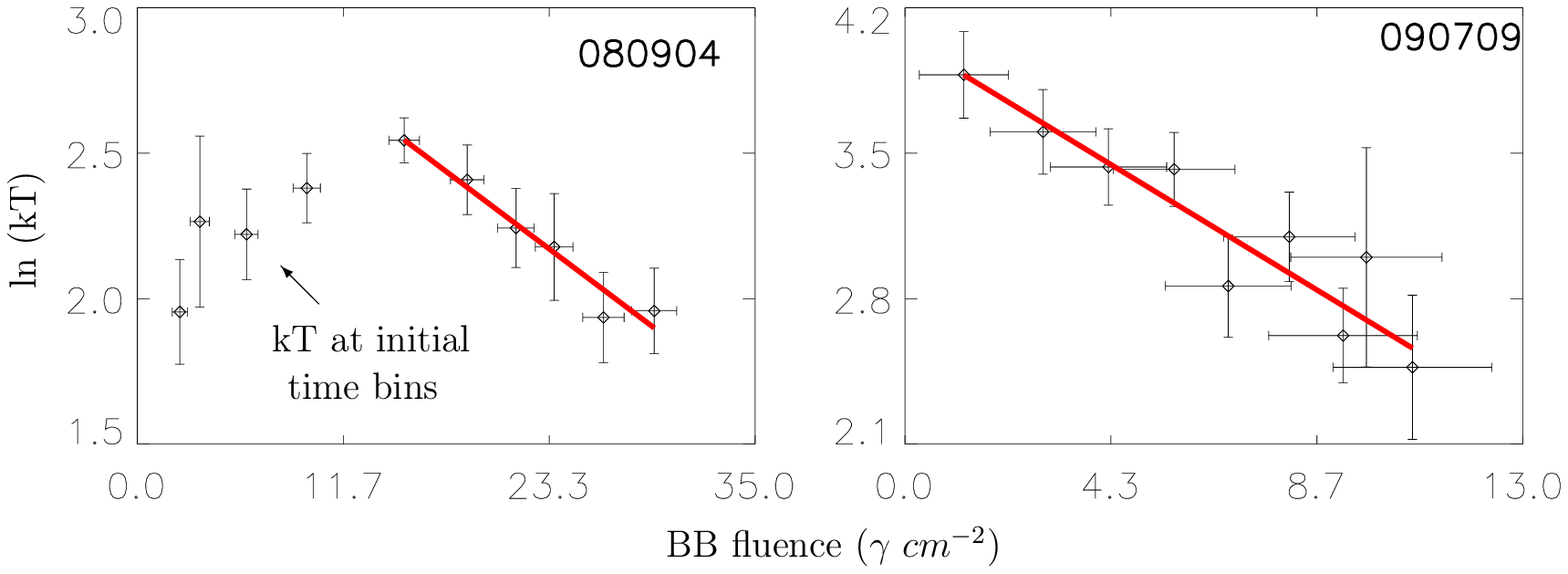}
   \includegraphics[width=3.2 in]{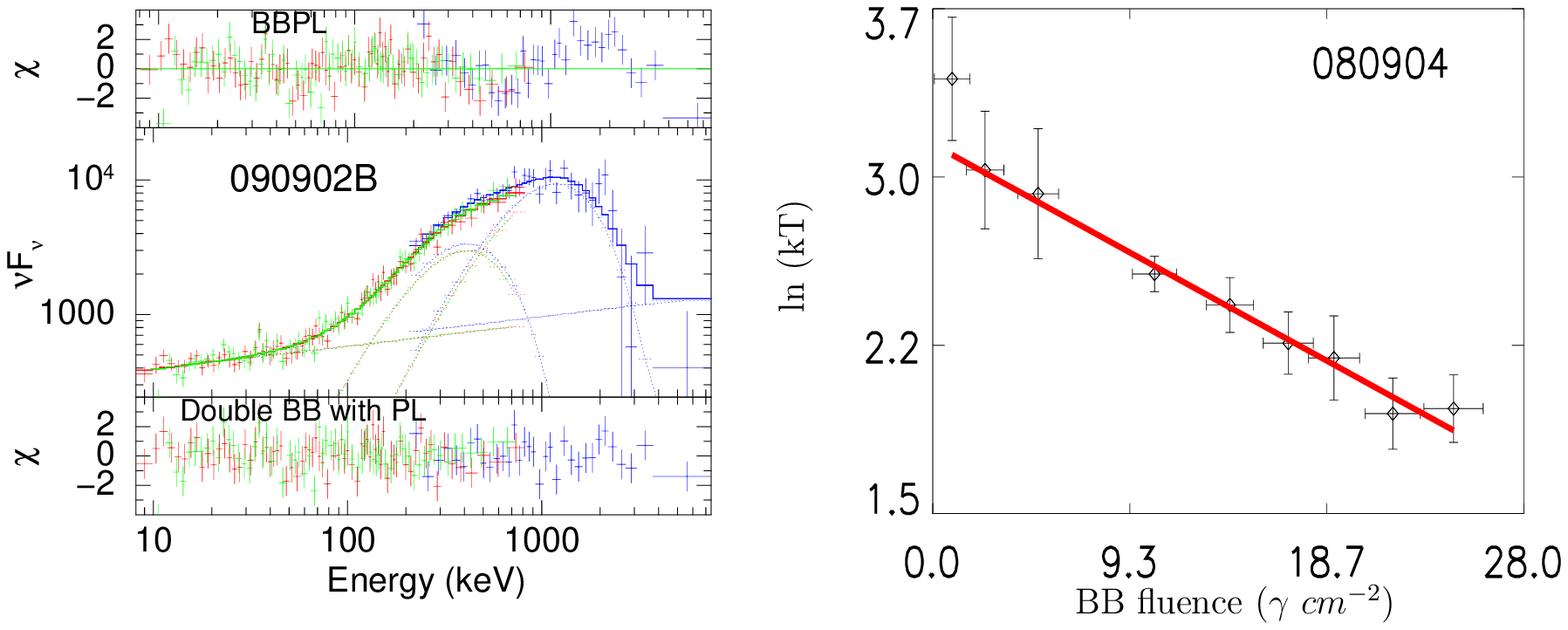}  
\caption{Variation of BB temperature (kT) with fluence for two representative GRBs. \emph{(upper left)}: kT
variation showing a break and \emph{(upper right)}: kT variation without a break.
\emph{(lower left)}: Double BB with PL, rather than BBPL, clearly fits the time-resolved spectrum of GRB 090902B.
 \emph{(lower right)}:  kT variation of GRB 080904 after putting double BB at the initial time bins.}
\end{figure}

\subsection{Alternate spectral models}
The improved correlation raises the hope of using GRB pulses as standard candles, instead of the full GRBs. 
But, the model is an empirical one. In order to refine the model on the basis of a physical description, we 
critically re-examine the spectral description, and try to find an alternative spectral model and its evolution, which is consistent
with the data. As pulses are important in our analysis, we collect, from the catalogue of Fermi/ GBM GRBs (Nava et al. 2011),
a set of GRBs having single or at least separable double pulses. We put a threshold on the brightness (fluence $> 10^{-6}$ erg 
cm $^{-2}$) and duration ($\delta t\geq 15$ s) and collect 11 GRBs. We fit the time-resolved spectra of these GRBs with the Band's 
GRB model and the BBPL model. The reduced $\chi^2$ of these fits have mean values, $\chi_{red}^2=$ 1.01 $\pm$ 0.19
and 1.07 $\pm$ 0.23 respectively, which points to the fact that both the descriptions are consistent with the data.

In Figure 2 (upper panels), we have plotted the natural log of temperature (kT, in keV) of the BB with fluence, for
two representative bursts. We find that kT, like the peak energy ($E_{peak}$), falls exponentially with the fluence. 
This law is always applicable in the later part of all pulses, and in the initial parts
we find breaks of kT/ $E_{peak}$ evolution for some pulses.
In fact, assuming that the radiation is due to photospheric emission of a fireball, we expect such a break (see e.g., Ryde 2004)
if the saturation radius ($r_{s}$) comes after the phototspheric radius ($r_{ph}$). Although, the expected kT 
variation is either constant (standard fireball) or slowly decreasing (magnetar model), but never increasing.
This reverse variation is apparent for GRB 080904, for example (Figure 2, upper left panel). Hence we 
investigated whether 
there are multiple spectral components affecting a smooth variation. We note, however, in Figure 2 (upper right panel) 
that GRB 090709 (and half of the sample, not shown here), shows a kT decrement throughout. Some show constant kT till the break.

Ryde et al. (2010) showed for a specific GRB, namely GRB 090902B, that the instantaneous spectra were consistent with 
a multicolour BB, rather than a BB. In an attempt
to fit the time-resolved data of this GRB we find that Band and BBPL models give unacceptable fits at many times. The BBPL fit 
gives two bumps in the residual (Figure 2, lower left panel). If we introduce another BB, then the
composite model 
 with two BB and a power-law (2BBPL)
gives acceptable fits. The average values of $\chi_{red}^2$ for the Band, BBPL and 2BBPL models are 1.64$\pm$0.97, 
1.60$\pm$0.25 and 1.09$\pm$0.13, respectively. The temperature and normalization of these two BB go hand-in-hand for all time 
with a tight correlation (r = 0.978 and 0.826 respectively). We find similar results for two other GRBs, namely, GRB 080916C and GRB 090926A.
Hence, we conclude that generally speaking, GRB pulses are consistent with double BB with a power-law. As the temperatures of
the BBs decrease for single pulses, as shown by all the pulses in our sample, the temperature of the lower BB hits the instrumental 
sensitivity limit, and hence does not show up in the falling part. But in the beginning 
part of the pulses, they might have an effect,
even if we find an acceptable $\chi_{red}^2$. 
In Figure 2 (lower right panel),the effect of putting double BB in the initial bins of GRB 080904 is shown. The kT variation
of the higher BB is now smooth throughout and we do not need a break in the kT evolution. 
For some GRBs (not shown), though we find a break, 
the kT is constant till the break. Hence the emerging  picture is consistent with the standard models.

\section{Conclusions and Future Work}
We summarize our work as follows: (a) A simultaneous timing and spectral description of GRB pulses has been developed, 
which correctly predicts derived parameters, e.g., width, spectral lag. (b) One of the major outcome of our model is finding 
a better correlation of $E_{peak,0}$, compared to $E_{peak}$, with $E_{\gamma,iso}$. Hence we conclude that $E_{peak,0}$ is the correct
parameter to use for Amati-type correlation study. (c) Alternative spectral model, e.g., BBPL, is consistent with the data. The 
temperature of the BB falls exponentially with the running fluence similar to the $E_{peak}$ of Band model. (d) If we force double
BBs in the initial part of a GRB pulse, then the temperature variation is either smooth throughout or constant till the break. 
The source of these two blackbodies is speculative. For example, they might be coming from different regions of photosphere boosted 
by different multiples of the bulk Lorentz factor ($\Gamma$). Alternately, if they are thermal supernova photons boosted by a cannon 
ball (CB; Dado et al. 2007), ejected by the central engine, then one identifies one of the components as the thermal inverse Compton 
and the other might be the bremsstrahlung photons radiated by the electrons.

In future, a comprehensive physical model of prompt GRB pulses can be constructed from a detailed calculation of the origin
of the observed spectral behaviours. The current model is an empirical one.
Once a pulse model is determined, the same can be used for the constituent pulses of a complicated GRB with known redshift.
This raises an enormous hope of using GRB pulses as standard candles. The fact that each pulse can be used for this purpose,
gives an extra constraint on the derived redshift and other related parameters.

\end{document}